\documentstyle[preprint,aps,prb]{revtex}

\begin{document}

\draft

\title{Final state effects on superfluid $^{\bf 4}$He in the deep
       inelastic regime}

\author{F.Mazzanti$^1$, J.Boronat$^2$ and A.Polls$^1$}
\address{ $^1$ Departament d'Estructura i Constituents de la
Mat\`eria, \\
          Diagonal 645, Universitat de Barcelona, \\
          E-08028 Barcelona, Spain}

\address{$^2$Departament de F\'{\i}sica i Enginyeria Nuclear, Campus
Nord B4-B5 \\
         Universitat Polit\`ecnica de Catalunya, \\
         E-08028 Barcelona, Spain}


\maketitle

\begin{abstract}

  A study of Final State Effects (FSE) on the dynamic structure function
of superfluid $^4$He in the Gersch--Rodriguez formalism is presented.
The main ingredients needed in the calculation are the momentum
distribution and the semidiagonal two--body density matrix.
The influence of these ground state
quantities on the FSE is analyzed. A variational form of
$\rho_2$ is used, even though simpler forms turn out to give accurate
results if properly chosen.
Comparison to the experimental response at high momentum
transfer is performed.
The predicted response is quite sensitive to
slight variations on the value of the condensate fraction, the best
agreement with experiment being obtained with $n_0=0.082$.
Sum rules of the FSE broadening function are also derived and commented.
Finally, it is shown
that Gersch--Rodriguez theory produces results as accurate as those
coming from other more recent FSE theories.

\end{abstract}


\narrowtext

\section {Introduction}

  Deep inelastic neutron scattering (DINS) has been extensively applied to the
study of quantum fluids, since Hohenberg and Platzman's~\cite{platz1} proposal
of using DINS to determine the momentum distribution $n(k)$ of helium atoms in
superfluid $^4$He.  The determination of $n(k)$ in quantum liquids is a
challenging problem of fundamental interest
.\cite{sok1} In fact, the knowledge of $n(k)$ provides very useful
information
to understand basic properties of the quantum nature of these systems as the
Bose--Einstein condensation. At
the same time, the theoretical analysis of DINS probes and stimulates the
development of modern many--body techniques.
These issues have been the main motivations of a considerable
amount of measurements and theoretical work on liquid $^4$He and other quantum
liquids.~\cite{glyde1,Sokol,sos1,sos2,sok3,gersch1,gersch2,gersch3,
sears1,silver1,rinat,carraro1,glyde2,bel1,mazz}

The inelastic scattering of neutrons by
liquid $^4$He is described by the double differential cross section
\begin {equation}
\frac {d^2 \sigma}{d\Omega d\omega } = b^2 \frac {k_f}{k_i} S(q,\omega)
\  ,
\end{equation}
where $b$ is the scattering length, $k_i$ and $k_f$ are the initial and final
wave vectors of the scattered neutron , and $q$ and $\omega$ are the momentum
and energy transferred from the neutron to the sample.  The dynamics of the
sample is entirely contained in $S(q,\omega)$, the dynamic structure factor,
which is the Fourier transform of the density-density correlation
function.\cite{lov1}  At sufficiently high momentum transfer, the
scattering
is entirely due to single atoms and $S(q,\omega)$ can be accurately described
by the Impulse Approximation (IA),~\cite{platz1} provided that the
interatomic potential does not have an infinite repulsive
core. In this regime, the kinetic
energy of an atom recoiling from a neutron collision is much larger than the
potential energy due to the interaction with the neighboring atoms, so that
collisions of the former with other atoms can be neglected. The
IA predicts a simple relation between $S(q,\omega)$ and $n(k)$,
\begin{equation}
S_{IA}(q,\omega)=\frac{1}{(2\pi)^3 \rho}\int d{\bf k} \ n( k)
\delta \left ( \omega - \omega_R - \frac { {\bf k \cdot q}} {m} \right
)
\label{introa1}
\end{equation}
where $\omega_R= q^2/2m$ is the free atom recoil frequency, $m$ is the
mass of
the $^4$He atoms and $n(k)$ is the thermally averaged occupation probability of
the single particle state of momentum ${\bf k}$, which reduces to that
of the
ground state at $T=0$. The delta function in Eq.(\ref{introa1}) takes
care of
the momentum and energy conservation in the scattering event between the
neutron and
a single atom. Assuming $S(q,\omega)=S_{IA}(q,\omega)$, the momentum
distribution $n(k)$ can be extracted from Eq.~(\ref{introa1}) by simple
differentiation. Notice that, in the previous equation and henceforth,
$\hbar$ is set to 1.

In isotropic systems, where $n(k)$ depends only on the
modulus of ${\bf k}$, it is useful to introduce the Compton profile
\begin{equation}
J_{IA}(Y)=\frac{q}{m} S_{IA}(q,\omega).
\label{intro4}
\end{equation}
which is driven by a single variable
\begin{equation}
Y = \frac{m}{q} \left( \omega - \frac{q^2}{2m} \right ),
\label{introb1}
\end{equation}
and fulfills $Y$ scaling.~\cite{west}
If a finite fraction of atoms $n_0$ occupies the zero momentum state,
$J_{IA}(Y)$ presents a $\delta$ peak of strength $n_0$ at $Y=0$.  However, this
expected signature of the condensate, is not observed in experiments performed
at momentum transfer as high as $23$\AA$^{-1}$,~\cite{Sokol} because
the IA spectrum is
broadened by both final state effects (FSE) and instrumental resolution effects
(IRE).  Hence, the theoretical interpretation of the experimental data requires
not only the knowledge of $n(k)$, but also an accurate description of
both the dynamics which determines FSE and the instrumental
broadening function.~\cite{sos1}

Several methods to account for FSE have been proposed.
\cite{gersch1,gersch2,silver1,rinat,carraro1,glyde2}
Among them, we will focus on the so--called convolutive theories, in
which
\begin{equation}
S(q,\omega)=\int _{-\infty}^{\infty} d\omega' S_{IA}(q,\omega') R(q,\omega
-\omega'),
\label{intro5}
\end{equation}
where $R(q,\omega)$ is the FSE broadening function.

After the first attempt~\cite {platz1} to approximate $R(q,\omega)$ by a
Lorentzian of width proportional to the $^4$He-$^4$He cross section, Gersch et
al.~\cite{gersch1} expressed the response function $S(q,\omega)$ in a
$1/q$
series expansion, whose coefficients are given by integrals of many-body
correlation functions averaged on the ground state of the system. In this
approach, the response when $q \rightarrow \infty$ is given by the first term
of the expansion of the incoherent part of $S(q,\omega)$, which turns to
be
exactly the IA.  However, the theory could not deal with realistic interatomic
potentials presenting a strong repulsion at short distances. To overcome
this problem, Gersch and Rodriguez~\cite{gersch2} proposed a cumulant
expansion of $S(q,t)$
which provides an adequate frame for calculating the response function at high
momentum transfer. The full calculation is impractical, but with some
approximations based on physical grounds, $S(q,\omega)$ can be expressed in
terms of the one- and  the semidiagonal two--body density matrices, and
the two
body interaction.  At the time the theory was proposed the numerical
application was made with a very simple approximation of the two-body density
matrix that resulted in an overestimation of the response at the
quasielastic peak.~\cite{gersch3}

The main purpose of the present work is to revisit Gersch--Rodriguez
theory,
and show that using a realistic two-body density matrix one gets a
$S(q,\omega)$ in good agreement with both experimental data and more
recent theories of FSE.~\cite{silver1,carraro1}

In the next section, a review of the theory is presented. Section III is
devoted to the discussion of the results and their comparison with the
experimental data. A sum rules analysis of $R(q,\omega)$ is presented in
section IV. In Section V our results are compared with other FSE
theories, and
finally section VI summarizes the main conclusions of the work.

\section {Gersch--Rodriguez Theory of FSE}

In the Gersch--Rodriguez theory,~\cite{gersch2} the density--density
correlation factor
$S(q,t)$  is expressed as the product of the IA and the FSE correcting function
by means of a cumulant expansion. The n--th order cumulant
accounts for the correlations among the struck atom and clusters of $n$
particles in the background.  In the high momentum transfer limit,
those terms with $n=1$ carry the most significant corrections. At this
level, the FSE broadening function can be expressed as a function of the
interatomic potential and the one- and two--body density matrices.

  The starting point in Gersch--Rodriguez theory is the time
representation of the response
\begin{equation}
 NS(q,t) = \sum_{j,l}
 <e^{-i{\bf q}{\bf r}_{l}}e^{iHt}e^{i{\bf q}{\bf r}_{j}}e^{-iHt}> =
 \sum_{j,l} <e^{i{\bf q}({\bf r}_{j}-{\bf r}_{l})}
 e^{-i{\bf q}{\bf r}_{j}}
 e^{iHt}e^{i{\bf q}{\bf r}_{j}}e^{-iHt}>  \ ,
\label{gsb1}
\end{equation}
which can be brought to the following form
\begin{equation}
 NS(q,t) = e^{i\omega_q t} \sum_{j,l} <e^{i{\bf q}({\bf r}_{j}-{\bf r}_{l})} \:
 e^{iL_{j}t} \: T \: \exp \left[ i\int_{0}^{t} dt' \: H({\bf r}_j -
 {\bf v}_{q}t')) \right] e^{-iHt}> \ ,
\label{gse1}
\end{equation}
where $T$ is the time--ordering operator and $H({\bf r}_j - {\bf v}_q
t')$ is the actual Hamiltonian
of the system where the position coordinate of particle $j$ has been
shifted by an amount ${\bf v}_q t'$. As the interatomic potential
considered is velocity--independent, one can write
\begin{equation}
   H({\bf r}_j - {\bf v}_q t') = H + U_j({\bf v}_q t') \ ,
\label{gsg1}
\end{equation}
with
\begin{equation}
   H=\sum_{j} \frac{p_{j}^{2}}{2m} + \sum_{i<j} V(r_{ij})
\label{gsg1b}
\end{equation}
and
\begin{eqnarray}
  U_{j}({\bf v}_q t') & = & \sum_{m \neq j} U_{j,m}(v_q t')   \nonumber   \\
  U_{j,m}(v_q t') & = & \left[ V({\bf r}_j -
  {\bf v}_q t', {\bf r}_m) - V({\bf r}_j, {\bf r}_m) \right]   \ ,
\label{gsg1c}
\end{eqnarray}
where $v_q = q/m$ and $\omega_q = q^2/2m$.

  The incoherent part of the response, which is defined by taking
particles labeled $j$ and $l$ in Eq.(\ref{gse1}) to be the same, is the
only contribution at large $q$. In this limit,
$S(q,t)$ may be written in the following way
\begin{equation}
 S(q,t) = e^{i\omega_q t} < e^{i{\bf v}_q t {\bf p}_1} e^{iHt}
 \: T \exp \left[ i\int_{0}^{t} dt'
 \sum_{m \neq 1} \hat{U}_{1,m}({\bf v}_q t') \right]  e^{-iHt} >  \ ,
\label{gsj1}
\end{equation}
where $\hat U({\bf v}_q t')$ is the previously defined potential operator but
with the position operators evaluated at time $t'$ rather than at $t=0$.
Notice that expression~(\ref{gsj1}) is as
hard to evaluate as the original $S(q,t)$.  An exact treatment would require
the knowledge of the time evolution of the whole system, so different
approximations should be made in order to deal with this last relation.

  Gersch and Rodriguez~\cite{gersch2} performed
a cumulant expansion of the ground state expectation value of
Eq.(\ref{gsj1}).  The expansion contains an infinite
number of terms, and allows for the
factorization of the IA from the total response
\begin{equation}
 S(q,t) = S_{IA}(q,t) \, R(q,t)    \ ,
                                                      \label{gsl3}
\end{equation}
$R(q,t)$ being the FSE correcting function given by
\begin{equation}
R(q,t) = \exp \left[ - \frac{1}{<e^{it{\bf v}_q {\bf p}_1}>}
\sum_{m \neq 1} <e^{it{\bf v}_q {\bf p}_1}
\left[ 1 - T \exp \left\{i\int_{0}^{t} dt'\hat
{U}_{1,m} ({\bf v}_q t') \right\} \right] > + ... \right]      \ .
\label{gsl4}
\end{equation}

 Up to this point, the result is exact because it
is nothing more than a rearrangement of the different terms entering in
$S(q,t)$. The first problem in the calculation of Eq.(\ref{gsl4})
is associated to the infinite number of terms appearing in
the exponential. Such a difficulty can be skipped if one
looks for the underlying
physics contained in each term: the contribution of
the $n$--th order cumulant to $S(q,t)$
accounts for the correlations between $n$-particle clusters during their
interactions with the struck atom. One may expect that the first significant
correction to the IA is produced by the multiple scattering of the struck
particle with the atoms of the media, considering them
independently
of each other. This corresponds to a truncation of the series beyond
the first order cumulant.

  The second problem lies on the evaluation of the
time--dependence appearing in the particle coordinates of $\hat
U_{1,m}({\bf v}_q t')$. In the large $q$ limit,
the displacement of the struck particle
is much larger
than the average movement of the background atoms. Thus, one can discard
the time dependence of ${\bf r}(t)$ in
$\hat{U}_{1,m}$. This is a safe
procedure as, even though the inclusion of such a time dependence
avoids hard--core collisions between the struck particle and other target
atoms, the contribution to $R(q,t)$ coming from those situations
vanishes due to rapid oscillations in the imaginary exponential of
Eq.(\ref{gsl4}).
Therefore, one can write~\cite{gersch2}
\begin{equation}
 R(q,t) = \exp \left[ - \frac{1}{\rho_1 (v_q t)} \int d{\bf r}
 \rho_2({\bf r},0;{\bf r} + {\bf v}_q t,0) \left[ 1 - \exp \left\{ i
 \int_{0}^{t} dt' \left( V({\bf r} + {\bf v}_q (t-t')) - V({\bf r} +
 {\bf v}_q t) \right) \right\} \right] \right] ,
                                               \label{gsl6}
\end{equation}
where $\rho_1$ and $\rho_2$ are the one--body and
semidiagonal two--body density matrices~\cite{clarkrist} of the
system, respectively. $R(q,t)$ is a complex function, but
its Fourier transform is real because its  real
part is even and its imaginary part odd under the change $t \to -t$.

  Equation~(\ref{gsl3}) predicts $S(q,t)$ as the
product of $S_{IA}(q,t)$ and $R(q,t)$, and therefore $S(q,\omega)$
is the convolution of $S_{IA}(q,\omega)$ and $R(q,\omega)$
\begin{equation}
 S(q,\omega) = \int_{-\infty}^{\infty} d\omega' S_{IA}(q,\omega') R(q,
 \omega - \omega') \ .
                                                   \label{gsl7}
\end{equation}

  In the particular case of liquid $^4$He, the momentum
distribution $n(k)$ may be written as
\begin{equation}
 n(k) = (2\pi)^3\rho n_0 \: \delta({\bf k}) + \tilde{n}(k)  \ ,
\label{gsl8}
\end{equation}
where $n_0$ is the condensate fraction value and
$\tilde{n}(k)$ stands for the occupation of non--zero momentum states.
Consequently, $S_{IA}(q,\omega)$ is split in two parts
\begin{equation}
 S_{IA}(q,\omega) = n_0 \delta \left( \omega - \frac{q^2}{2m} \right)
 + \frac{m}{4\pi^2 \rho q} \int_{\mid \frac{m\omega}{q} - \frac{q}{2}
  \mid}^{\infty} \, k \: n(k) \, dk
 =  n_0 \delta \left( \omega - \frac{q^2}{2m} \right) +
 \tilde{S}_{IA}(q,\omega)  \ .
                                                    \label{gsm3}
\end{equation}
where the first term on the rhs is the condensate response which appears as
a delta peak of strength $n_0$ located at the quasielastic energy, and
$\tilde{S}_{IA}(q,\omega)$ is the non--condensate contribution of $n(k)$
to the IA. Introducing the West variable $Y=m\omega/q-q/2$, $S_{IA}(q,\omega)$
can be expressed in terms of the Compton Profile
\begin{equation}
 \frac{q}{m}S_{IA}(q,\omega) \equiv J_{IA}(Y) = n_0\delta(Y) +
 \frac{1}{4\pi^2 \rho} \int_{\mid Y \mid}^{\infty}
 k\, n(k) \ dk ,
                                                     \label{gsm4}
\end{equation}
which scales in $Y$

Moreover, at high $q$  the response is usually expressed in
terms of $Y$ through the relation
\begin{equation}
 J(q,Y) = \frac{q}{m} S(q,\omega)       \ ,
                                                     \label{gsm5}
\end{equation}
and thus, Eq.(\ref{gsl7}) is transformed into
\begin{equation}
 J(q,Y) = \int_{-\infty}^{\infty} \, dY' J_{IA}(Y') R(q,Y-Y') =
 n_0 R(q,Y) + \int_{-\infty}^{\infty} dY' \tilde{J}_{IA}(Y') \,
R(q,Y-Y') \ ,
                                                     \label{gsm6}
\end{equation}
where
\begin{equation}
   R(q,Y) = \frac{q}{m}\, R(q,\omega)    \ .
\label{gsm6b}
\end{equation}

\section {Numerical Results}

In this section, we present results for the FSE correcting function
$R(q,Y)$ and the response function $J(q,Y)$ calculated in the framework of the
Gersch--Rodriguez formalism.
The input density matrices $\rho_1(r)$ and $\rho_2({\bf r_1}, {\bf r_2}; {\bf
r_1'},{\bf r_2})$ used to calculate $J_{IA}(Y)$ and $R(q,Y)$ have been obtained
in the framework of the HNC theory~\cite{fantoni,manou,clarkrist}
from a variational many body wave function containing two- and three--body
correlations.~\cite{boro2} The variational minimization has been
performed for the HFDHE2
Aziz potential~\cite{Aziz} at the experimental equilibrium density
($\rho=0.365
\sigma^{-3}$, $\sigma=2.556$\AA). The ground state description obtained with
this wave function is in good agreement with recent
Green's function Monte Carlo
calculations.~\cite{boro,pan}  The discussion is separated in two
parts, the first one being devoted to the study of both $R(q,Y)$ and
$J(q,Y)$ and
their comparison to experimental data, and the second one to the
analysis of the
dependence of these functions on the different approximations used in
the variational description of the ground state wave function.

The actual calculation of the FSE broadening function is initially performed in
time representation (\ref{gsl6}). $R(q,x)$ is a complex quantity which can be
written in the following way:
\begin{equation}
R(q,x)  =  e^{\phi(q,x)} \left[ \cos(\psi(q,x)) + i \sin(\psi(q,x))
\right]
\label{noseque}
\end{equation}
with
\begin{eqnarray}
 \phi(q,x) & = & -\frac{2}{\rho_1(x)} \int d{\bf r}
\rho_2({\bf r}, 0;{\bf r}+{\bf x}) \sin^2 \left[ \frac{1}{2 v_q}
\int_0^x du \left\{ V({\bf r}+{\bf x}-{\bf u}) - V({\bf r}+{\bf x})
\right\} \right]     \nonumber \\
\psi(q,x) & = & \frac{1}{\rho_1(x)} \int d{\bf r}
\rho_2({\bf r}, 0;{\bf r}+{\bf x}) \sin \left[ \frac{1}{ v_q}
\int_0^x du \left\{ V({\bf r}+{\bf x}-{\bf u}) - V({\bf r}+{\bf x})
\right\}  \right]  \ ,
\label{numres-a1}
\end{eqnarray}
$x$ being $v_q t$.  As can be seen form Eq.(\ref{numres-a1}),
$\phi(q,x)$ and $\psi(q,x)$ are even and odd functions of $x$,
respectively.
Therefore, the real and imaginary parts of $R(q,x)$ are
respectively even and odd
under the change $x \to -x$, and consequently $R(q,Y)$ is  real.
Even if the potential becomes very repulsive at short distances, as
is the case of the Aziz potential, Eq.(\ref{gsl6}) gives an $R(q,Y)$
which does not diverge.

The real and imaginary parts of $R(q,x)$ are shown in
Fig. 1 for $q=23.1$\AA$^{-1}$.
In the relevant range of $x$, Re$R(q,x)$ has a dominant decreasing
behavior.
The Re$R(q,x)$ and Im$R(q,x)$ are related to the symmetric and
antisymmetric
components of $R(q,Y)$, respectively. As the imaginary part is much smaller
than the real part, $R(q,Y)$ is mostly symmetric around $Y=0$.
 In Fig. 2, we show $\psi(q,x)$ and
$\phi(q,x)$ at $q=23.1$\AA$^{-1}$. $\phi(q,x)$ is a negative and a
monotonously decreasing function of
$x$, causing both the real and the
imaginary parts of $R(q,x)$ tend to zero when $x \to
\infty$~(\ref{noseque}).

In Fig. 3, we show $R(q,Y)$ at two different values of $q$, 23.1
\AA$^{-1}$ and 15.0 \AA$^{-1}$. The
main trends of $R(q,Y)$ in all FSE convolution theories are the same: a
dominant central peak and small oscillating tails which vanish as $|Y|$
increases. As one can see, the shape of $R(q,Y)$ smoothly changes
with $q$, this variation being reflected in an overall redistribution of the
strength between the main peak and the wings. When $q$ increases, the peak
appears higher and narrower pointing to the tendency of $R(q,Y)$ to become a
delta distribution in the limit $q \rightarrow \infty$.

The existence of a finite condensate fraction $n_0$ in superfluid $^4$He
plays an important role in the FSE corrections, as is reflected in
Fig. 4 where the broadening of the
condensate and non--condensate parts of $J_{IA}(Y)$ are separately shown.
The small differences between $\tilde{J}_{IA}(Y)$
(dotted line) and the convolution of $\tilde{J}_{IA}(Y)$ with
$R(q,Y)$ (long--dashed line) reveal small FSE on the
non--condensate part of the response at high $q$.
In contrast,
the broadening of the condensate term (short--dashed line), i.e., the
convolution product of $R(q,Y)$ and $n_0 \, \delta(Y)$, contributes to
$J(q,Y)$ as $n_0 \, R(q,Y)$ which is a function with an appreciable width and
height. The inclusion of the latter term produces a total $J(q,Y)$
(solid line) which manifests a sizeable departure from the IA
prediction. Therefore, FSE corrections in superfluid $^4$He appear
to be relevant even at so high $q$'s.~\cite{silver1}

A direct comparison between theoretical and experimental dynamic
structure factors is not possible due to the presence of instrumental
resolution effects (IRE) in the experimental data acquisition process.
It would be desirable, from a theoretical viewpoint, to remove the IRE
inherent to the measured response, especially at high $q$ where they
become larger. However, the latter is an ill--posed problem due to the
statistical noise of the data, and thus the only way to compare theory
and experiment is by convoluting the theoretical $J(q,Y)$ with an
instrumental resolution function $I(q,Y)$. At present, $I(q,Y)$ is
obtained from a Monte Carlo simulation of the experimental setup,
and in contrast to earlier models used in neutron scattering analysis,
it is neither Gaussian nor symmetric around $Y=0$, and is comparable in
width and height to $R(q,Y)$ at those momenta.~\cite{Sokol} The
influence of $I(q,Y)$ in the response is sketched in Fig. 5 for $q=23.1$
\AA$^{-1}$.  As one can see, the introduction of the IRE in the
response (solid line) appreciably modifies $J(q,Y)$ (dashed line). The
most important effect of $I(q,Y)$ is to quench the central peak
reducing the effects of the FSE correction on $J_{IA}(Y)$,
whereas the tails remain almost unchanged.

In Fig. 6, we present results of $J(q,Y)$ broadened by the IRE
at different values of $q$
in comparison with inelastic scattering data at $T=0.34$ K from
Ref.~\onlinecite{Sokol}. There is an overall agreement between the
predicted and the
observed scattering data, the quality of the Gersch--Rodriguez theory being
comparable to results provided by other theories~\cite{silver1,carraro1} (see
also section V).
It is worth to notice that
all FSE theories are stressed when applied to intermediate $q$
values. This is also apparent in our results, as one can see for the
lowest $q$ value reported in Fig. 6. Thus, whereas the experimental
peak shifts its location to a small negative $Y$ value, the
theoretical one
is shifted to so small positive values of $Y$ that it is
not appreciable in the figure.

  The most relevant quantity in
the calculation of $J(q,Y)$ is the momentum distribution $n(k)$ which
completely determines the Compton profile $J_{IA}(Y)$. The influence of
$n(k)$ in $J(q,Y)$ is shown in Fig. 7. The dashed and solid lines
correspond to a Jastrow $n(k)$ ($n_{J}(k)$) and a
Jastrow plus triplet $n(k)$ ($n_{JT}(k)$), respectively. The condensate
fraction predicted by the two approximations are slightly different,
$n_0^J=0.091$ and $n_0^{JT}=0.082$. This reduction of $n_0$, produces a small
decrease of strength in the peak of $J(q,Y)$ bringing our
theoretical prediction closer to the experiment.
A basic ingredient in the calculation of $R(q,Y)$ is the semidiagonal
two--body density matrix,
 which in the framework of the HNC theory is given by~\cite{clarkrist}
\begin{equation}
\rho_2({\bf r}_1, {\bf r}_2; {\bf r'}_1, {\bf r}_2) = \rho \rho_1(r_{11'})
g_{wd}(r_{12}) g_{wd}(r_{1'2}) \exp \left[ A({\bf r}_1, {\bf r}_2; {\bf r'}_1)
\right]
\label{otraeq}
\end{equation}
where $\rho_1(r_{11'})$ is the one--body density matrix, $g_{wd}(r)$ is an
auxiliary two--body radial distribution function and $A({\bf r}_1, {\bf r}_2;
{\bf r'}_1)$ is the sum of the Abe diagrams. Notice that the
structure of $\rho_2$ allows for the exact cancellation of $\rho_1$ in
Eq.~(\ref{gsl6}). As the explicit dependence of $\rho_2$ in $n_0$ is introduced
in $\rho_1$,
the influence of $n_0$ in $R(q,Y)$ is almost negligible.
We have verified that the inclusion of three--body
correlations does not appreciable modify the structure of $R(q,Y)$.
 Consequently, three--body correlations can be omitted in the
calculation of $R(q,Y)$. In a further step, we have also studied the influence
of the Abe diagrams using a Jastrow wave function.
As  is well known, it is not
possible to calculate $ A({\bf r}_1, {\bf r}_2, {\bf r}_1')$ exactly
but a good estimation of its contribution
can be obtained through the scaling approximation. \cite{Scaling}  The
inclusion of the Abe diagrams in Eq. (\ref{otraeq}) using the scaling
approximation produce negligible effects in the final form
of $R(q,Y)$. In fact, the Abe terms, which quickly vanish when the
inter-particle separation increases, only modify the structure of
$\rho_2$ when coordinates 1, 1$^{\prime}$ and 2 are very close to each
other. These small changes in $\rho_2$ are then suppressed when
integrated to obtain $R(q,t)$. Furthermore, one can slightly change
the functions $g_{wd}(r)$ and no influences in $R(q,Y)$ are
observed. This fact, which will be explicitly commented in Sec.  V,
points to the relevance of the functional decomposition of $\rho_2$
rather than the exact form of the functions entering in
it.

\section {Sum Rules}

  In this section we study the sum rules satisfied by
the Gersch--Rodriguez FSE broadening function $R(q,Y)$.
{}From the relation
\begin{equation}
 J(q,Y) = \int_{-\infty}^{\infty} dY' J_{IA}(Y') R(q,Y-Y')  \ ,
\label{sra1}
\end{equation}
and the first sum rules of both $J_{IA}(Y)$ and the incoherent
part of $J(q,Y)$, an equivalent
set of $Y$--weighted integrals for $R(q,Y)$ can be
derived.~\cite{silver1}  Notice
that equation (\ref{sra1}) can be taken as a possible definition of $R(q,Y)$
provided that $q$ is large enough for the coherent part of $J(q,Y)$ to be
negligible.  These sum rules are model independent, and so any suitable
convolutive FSE broadening function must fulfill them. The first sum rules of
$R(q,Y)$ are
\begin{eqnarray}
 & & m_{0}^{R}(q) = \int_{-\infty}^{\infty} dY R(q,Y) = 1 \nonumber \\
 & & m_{1}^{R}(q) = \int_{-\infty}^{\infty} dY Y R(q,Y) = 0 \nonumber \\
 & & m_{2}^{R}(q) = \int_{-\infty}^{\infty} dY Y^{2} R(q,Y) = 0
\nonumber \\
 m_{3}^{R}(q) & = & \int_{-\infty}^{\infty}
dY Y^{3} R(q,Y) = \frac{m}{2 q^3} \rho \int d{\bf r} g(r) ({\bf q}
\cdot {\bf \nabla})^2 V(r)  \ .
\label{sra2}
\end{eqnarray}

  As we are only considering the incoherent
part of the response, $m_0^R(q)$ is 1 at any $q$.  Both the first and
second moments of $R(q,Y)$ vanish
because the Impulse Approximation exactly fulfills the incoherent sum rules.
Finally, the third moment of $R(q,Y)$ is expressed in terms of the
two--body radial distribution function $g(r)$ and the interatomic
potential $V(r)$, which are not included in $J_{IA}(Y)$.

  Relations~(\ref{sra2}) are exact and partially define the
behavior of
$R(q,Y)$. Therefore, one can use them to check the accuracy of $R(q,Y)$
calculated using different approximations. In the Gersch--Rodriguez
theory, the sum rules analysis can be analytically performed.  In
fact, expressions for the sum rules can be easily derived from the
time derivatives of $R(q,t)$ at $t=0$
\begin{equation}
m_{k}^{R}(q) = \frac{1}{i^k v_q^k} \left. \frac{d^{k}}{dt^{k}}
R(q,t) \right|_{t=0}          \ .
\label{sra3}
\end{equation}

  Performing a McLaurin expansion of $R(q,t)$, Eq.~(\ref{gsl6}), the
different coefficients of the series are directly related to the
$Y$--weighted sum rules. In this way, one obtains the relations
\begin{eqnarray}
 m_{0}^{R,GR}(q) & = & 1 \nonumber   \\
 m_{1}^{R,GR}(q) & = & 0 \nonumber   \\
 m_{2}^{R,GR}(q) & = & 0 \nonumber   \\
 m_{3}^{R,GR}(q) =  \frac{2m}{q^3 \rho} \int d{\bf r}
 \rho_2({\bf r}, 0;{\bf r}) ({\bf q}
 \cdot {\bf \nabla})^2 V(r) & + &
  \frac{3m}{q^3 \rho} \int d{\bf r} \left[ ({\bf q}\cdot\nabla) V(r)
 \right] \left[ ({\bf q}\cdot\nabla_{{\bf x}}) \rho_2({\bf r},0;{\bf x})
 \right]_{{\bf x} = {\bf r}}
\label{srb1}
\end{eqnarray}
where $m_{k}^{R,GR}(q)$ stand for the $Y$--weighted integral of the FSE
function in Gersch--Rodriguez theory.
Integrating by parts the second term of $m_{3}^{R,GR}(q)$,
and taking into account general symmetry properties of
$\rho_2$, one can express $m_3^{R,GR}(q)$ in the following way
\begin{equation}
m_3^{R,GR} (q) = \frac{m}{2q^3 \rho} \int d{\bf r}
\rho_2({\bf r}, 0; {\bf r}) ({\bf q} \cdot \nabla)^2 V(r) \ .
\label{srna1}
\end{equation}
  As the diagonal part of $\rho_2$ is $\rho^2 g(r)$, the analytic
expression of $m_3^R(q)$ is recovered. Therefore, the zero, first, second
and third moments of $R(q,Y)$ are exactly fulfilled in the
Gersch--Rodriguez theory.

  Nevertheless, the exact $\rho_2$ is not known, and the use of an
approximation can produce numerical differences between
Eqs.~(\ref{srna1}) and (\ref{sra2}). In fact,
we have checked that the inclusion
of the Abe terms in the variational $\rho_2$ defined in
Eq.~(\ref{otraeq}) is crucial in reproducing $g(r)$ in its
diagonal part, and consequently $m_3^R(q)$.

\section{Comparison with other FSE theories}

  FSE theories can be classified in
different groups depending on the way they incorporate the corrections to the
IA. The two most important groups are, on one hand, convolutive theories in
which the total response is expressed as a convolution of $J_{IA}(Y)$ and
$R(q,Y)$ and, on the other, additive theories where the leading  FSE
corrections are summed up to the IA.  Examples of theories belonging to the
first class are those of Silver \cite{silver1} or  Carraro and Koonin.
\cite{carraro1} An example of additive theory is that originally derived
by Gersch, Rodriguez and Smith,~\cite{gersch1} which was next
generalized by Rinat~\cite{rinat}
to treat also hard core potentials.

  Gersch--Rodriguez formalism was the first in predicting
convolutive FSE corrections. Silver's and Carraro and Koonin's theories
appeared some years after.
In this section, we present a comparison
between their results and our predictions obtained in the framework
of the Gersch--Rodriguez theory.

  In the Gersch--Rodriguez theory $R(q,Y)$ is
formulated in terms of the semi--diagonal two--body density matrix of the
system. In the present work, a variational {\em ansatz} for this
quantity has
been employed and discussed, but at the time the formalism was developed only a
qualitative description of $\rho_2$ was available. This led the original
authors to use a form of $\rho_2$ based on a Hartree--Fock
approximation and the Schwartz inequality~\cite{gersch1}
\begin{equation}
 \rho_2({\bf r}_1,{\bf r}_2;{\bf r'}_1,{\bf r}_2) = \rho
 \rho_1(r_{11'}) \sqrt{ g(r_{12}) g(r_{1'2})} \ ,
\label{compa1}
\end{equation}
$\rho_1(r)$ being the one--body density matrix and $g(r)$ the
two--body radial distribution function.
At that time, detailed microscopic calculations of $g(r)$ were not available,
so they had to approximate it. The form selected for
the radial distribution function was simply a step function
\begin{equation}
 g(r) = \theta (r - r_0) \ ,
\label{compa2}
\end{equation}
with  a parameter $r_0$ to mimic the radius of the hole of $g(r)$. Originally,
$r_0$ was taken as a fitting parameter.  However, theoretical arguments brought
them to fix its value to $r_0=2.5$\AA.~\cite{gersch3} With this
prescription,
Gersch and Rodriguez predicted a $J(q,Y)$ that visibly overestimates the
measured strength of the response around its maximum. This failure was
later
discussed and partially attributed to a somewhat excessively simplified
approximation to the problem.~\cite{silver1} Nevertheless, this
discrepancy
seems to be eliminated by  choosing a different value of $r_0$.  In order to
show this feature, several calculations using Eqs.(\ref{compa1}) and
(\ref{compa2}) with different values of $r_0$ have been performed.
In Fig. 8,
results for $R(q,Y)$ with $r_0$ equal to 2.0\AA, 2.1\AA and 2.2\AA are
depicted and compared to $R(q,Y)$ computed with the variational $\rho_2$.  Even
though the behavior of the tails of $R(q,Y)$ in the Gersch--Rodriguez
approximation of $\rho_2$ is different from the one of $R(q,Y)$ with
the
variational two--body density matrix, the height and width of both
peaks are in
good agreement for a value of $r_0$ laying between 2.1 and 2.2\AA.  Then, a
proper choice of $r_0$ in the simple Gersch--Rodriguez model for $\rho_2$
produces accurate results, provided that the height and width of the central
peak are the most important features of the FSE broadening function.

We have compared our results for $R(q,Y)$ and $J(q,Y)$ with those
obtained by Silver~\cite{silver1} and Carraro and
Koonin.~\cite{carraro1}
Figures 9, 10 and 11 show
$R(q,Y)$ and $J(q,Y)$ in Gersch--Rodriguez (GR), Silver (HCPT) and
Carraro and Koonin (CK) theories for three values of $q$,
23.1\AA$^{-1}$, 15.5\AA$^{-1}$ and 10.2\AA$^{-1}$.
The FSE function $R(q,Y)$ is slightly different in the three
theories, though both the height and width of the central peak
are quite similar. The tails of the FSE broadening function show a
different behavior, although they quickly vanish as $\mid Y \mid$ increases.
Despite of the discrepancies in $R(q,Y)$,
the predicted responses are nearly the same at $q=23.1$ \AA$^{-1}$ and
in good agreement with the experimental data.
As $q$ is lowered, the deficiencies of the FSE theories show up but
$J(q,Y)$ is still reasonably well described at $q=15.5$ \AA$^{-1}$. For
the lowest $q$ value, $q=10$ \AA$^{-1}$ (Fig. 11), the theoretical
responses move away from experiment,
and in particular do not present the small shift of the
peak to negative $Y$ values (see also Fig. 6). Then, even for
intermediate $q$ values,
the Gersch--Rodriguez theory reproduces the dynamic structure
function  as precisely as other existing theories for the FSE.

\section{Summary and Conclusions}

  In this work, Final State Effects on the density response of
superfluid $^4$He have been studied in the framework of the
Gersch--Rodriguez theory using a realistic description of
the ground state of the liquid. The response is predicted as the
convolution product of the Compton Profile $J_{IA}(Y)$ and the FSE
broadening function $R(q,Y)$.

  Two quantities describing the ground state of the system are needed.
The first one is the momentum distribution $n(k)$, which completely
determines  $J_{IA}(Y)$. The
second one is the semidiagonal two--body density matrix,
which  enters in the Gersch--Rodriguez form of $R(q,Y)$.

$J_{IA}(Y)$ has two terms, one corresponding to the
non--condensate part of $n(k)$ and another given by $n_0
\delta(Y)$. This splitting
produces, after convoluting with $R(q,Y)$, a total response which is
also the sum of two terms, corresponding to
the condensate and non--condensate contributions. The former is
linear in $n_0$ and mostly affects
$J(q,Y)$ around $Y=0$. The latter is much
less affected by FSE, although the effects are non--negligible.
We have verified that Gersch--Rodriguez theory gives accurate results
when proper forms for the one- and two--body density matrices are used.
A variational $\rho_2$ obtained in the HNC framework accurately
reproduces the experimental response at high $q$'s. Furthermore, we have
checked that the functional decomposition of $\rho_2$ is very
important in the calculation of $R(q,Y)$. Simple models conserving the
variational functional form can also produce a good estimation of the
response.

Our results
are comparable  to other calculations using  more recent convolutive
FSE theories. None of the theories correctly accounts for the
observed response when $q$ is lowered below about
10\AA$^{-1}$. Further improvements could arise when
higher order terms in the Gersch--Rodriguez cumulant expansion are
considered or the time dependence of the particle coordinates
is taken into account.

\acknowledgments
  The authors would like to thank H.R. Glyde and A.S. Rinat
for valuable suggestions and discussions. We are also indebted to
P.E. Sokol for many comments and for making his experimental data
available to us. This work was partially supported by DGICYT (Spain)
Grant No. PB92-0761. F. Mazzanti acknowledges his fellowship from the
Generalitat de Catalunya.

\begin{figure}
\caption{Real and imaginary parts of $R(q,x)$ at $q=23.1$\AA$^{-1}$.}
\end{figure}

\begin{figure}
\caption{Functions $\phi(q,x)$ and $\psi(q,x)$ at $q=23.1$\AA$^{-1}$.}
\end{figure}

\begin{figure}
\caption{$R(q,Y)$ at $q=23.1$ \AA$^{-1}$ (solid line) and
$q=15.0$ \AA$^{-1}$ (dashed line).}
\end{figure}

\begin{figure}
\caption{Different contributions to $J(q,Y)$ at $q=23.1$
\AA$^{-1}$.
Dotted line, non--condensate term of $J_{IA}(Y)$; long--dashed line,
non--condensate
term of $J_{IA}(Y)$ after the convolution with $R(q,Y)$;
short--dashed line, condensate contribution once broadened by FSE;
solid line, total response.}
\end{figure}

\begin{figure}
\caption{Effects of the different broadenings to the response at
$q=23.1$ \AA$^{-1}$. Dotted
line, non--condensate IA prediction; dashed line, IA broadened by FSE;
solid line, total $J(q,Y)$ including both FSE and IRE.}
\end{figure}

\begin{figure}
\caption{Comparison of the predicted $J(q,Y)$ at (a) $23.1$
\AA$^{-1}$, (b) 17.9\AA$^{-1}$, (c) 15.0\AA$^{-1}$, and (d)
10.2\AA$^{-1}$ with experimental data (points with error bars).}
\end{figure}

\begin{figure}
\caption{Detail of the central peak of the response at $q=23.1$
\AA$^{-1}$ as predicted using two different $n(k)$'s. The solid and
dashed lines correspond to $n_{JT}(k)$ and $n_J(k)$, respectively.
The points with errorbars are the experimental data.}
\end{figure}

\begin{figure}
\caption{Comparison of $R(q,Y)$ calculated using the
Gersch--Rodriguez $\rho_2$ with different values of $r_0$ (solid
lines) and the variational $\rho_2$ (dashed line).}
\end{figure}

\begin{figure}
\caption{Comparison between Gersch--Rodriguez, Silver, and Carraro and
Koonin results for both $R(q,Y)$ and $J(q,Y)$ at $q=23.1$ \AA$^{-1}$.}
\end{figure}

\begin{figure}
\caption{Comparison between Gersch--Rodriguez and Silver
results for both $R(q,Y)$ and $J(q,Y)$ at $q=15.0$ \AA$^{-1}$.}
\end{figure}

\begin{figure}
\caption{Comparison between Gersch--Rodriguez and Carraro and
Koonin results for both $R(q,Y)$ and $J(q,Y)$ at $q=10.2$ \AA$^{-1}$.}
\end{figure}

\end{document}